\documentclass[fleqn,usenatbib]{mnras}
\usepackage{newtxtext,newtxmath}
\usepackage[T1]{fontenc}
\DeclareRobustCommand{\VAN}[3]{#2}
\let\VANthebibliography\thebibliography
\def\thebibliography{\DeclareRobustCommand{\VAN}[3]{##3}\VANthebibliography}

\usepackage{graphicx}	
\usepackage{amsmath}	
\usepackage{xcolor}
\usepackage[version=4]{mhchem} 
\usepackage{booktabs,multirow}

\newcommand{\nbody}{$N$-body}
\newcommand{\hi}{H\scalebox{1.0}[0.8]{I}}
\newcommand{\pyccray}{\texttt{pyC$^2$Ray}}
\newcommand{\ccray}{\texttt{C$^2$Ray}}
\newcommand{\pkdgrav}{\texttt{Pkdgrav3}}
\newcommand{\lya}{Ly$\alpha$}
\defcitealias{hirling2024pyc}{Hirling, Bianco et al. 2024}
\newcommand{\xHIv}{$\langle x_\mathrm{HI}\rangle_\mathrm{v}$}
\newcommand{\xHIm}{$\langle x_\mathrm{HI}\rangle_\mathrm{m}$}

\newcommand{\RefereeReport}[1]{\textcolor{black}{#1}}
\newcommand{\RefereeReportTwo}[1]{\textcolor{black}{#1}}

\title[End stages of reionization]{The 21-cm signal during the end stages of reionization}

\author[S. K. Giri et al.]{Sambit K. Giri,$^{1}$\thanks{E-mail: sambit.giri@su.se}, Michele Bianco$^{2,3,5}$, Timoth\'ee Schaeffer$^{4}$,
Ilian T. Iliev$^{5}$,
Garrelt Mellema$^{6}$, 
and
\newauthor
Aurel Schneider$^{4}$
\\
$^{1}$ Nordita, KTH Royal Institute of Technology and Stockholm University, Hannes Alf\'vens v\"ag 12, SE-106 91 Stockholm, Sweden\\
$^{2}$ Laboratoire d’Astrophysique, Ecole Polytechnique Federale de Lausanne (EPFL), Observatoire de Sauverny, Versoix 1290, Switzerland\\
$^{3}$ Institute for Particle Physics and Astrophysics, ETH Zurich, Wolfgang-Pauli-Str 27, 8093 Zurich, Switzerland\\
$^{4}$ Department of Astrophysics, University of Zurich, Winterthurerstrasse 190, 8057 Zurich, Switzerland.\\
$^{5}$ Astronomy Centre, Department of Physics \& Astronomy, Pevensey III Building, University of Sussex, Falmer, Brighton, BN1 9QH, United Kingdom\\
$^{6}$ Department of Astronomy and Oskar Klein Centre, AlbaNova, Stockholm University, SE-10691 Stockholm, Sweden
}

\date{Accepted XXX. Received YYY; in original form ZZZ, Report Number NORDITA 2024-003}

\pubyear{2024}

\begin{document}
\label{firstpage}
\pagerange{\pageref{firstpage}--\pageref{lastpage}}
\maketitle

\begin{abstract}
During the epoch of reionization (EoR), the 21-cm signal allows direct observation of the neutral hydrogen (\hi{}) in the intergalactic medium (IGM). In the post-reionization era, this signal instead probes \hi{} in galaxies, which traces the dark matter density distribution.
With new numerical simulations, we investigated the end stages of reionization to elucidate the transition of our Universe into the post-reionization era. Our models are consistent with the latest high-redshift measurements, including ultraviolet (UV) luminosity functions \RefereeReport{up to redshift $\simeq$8}. Notably, these models consistently reproduced the evolution of the UV photon background, which is constrained from Lyman-$\alpha$ absorption spectra. We studied the dependence of this background on the nature of photon sinks in the IGM, requiring mean free path of UV photons to be $\sim$10 comoving-megaparsecs (cMpc) during the EoR that increases gradually with time during late stages ($z\lesssim 6$). Our models revealed that the reionization of the IGM transitioned from an \textit{inside-out} to an \textit{outside-in} process when the Universe is less than 0.01 per cent neutral. During this epoch, the 21-cm signal also shifted from probing predominantly the \hi{} in the IGM to that in galaxies. Furthermore, we identified a statistically significant number of large neutral islands (with sizes up to 40 cMpc) persisting until very late stages ($5 \lesssim z \lesssim 6$) that can imprint features in Lyman-$\alpha$ absorption spectra and also produce a knee-like feature in the 21-cm power spectrum. 
\end{abstract}



\section{Introduction}
Recent cutting-edge observational facilities, such as the James Webb Space Telescope (JWST), have substantially improved our capabilities to explore the high-redshift ($z\gtrsim 5$) Universe. In recent years, the dataset of early galaxies and quasars has been quickly growing \citep[e.g.][]{oesch2016remarkably,atek2018extreme,atek2023revealing,bakx2023deep,robertson2023discovery,bunker2023jades}. These observations are enhancing our understanding of both cosmology and astrophysics during the epoch of reionization (EoR), which is the period when early photon sources formed and ionized the gas in the intergalactic medium (IGM)\citep[e.g.][]{sun2023bursty,Schaeffer2023beorn,Hassan2023UV,dayal2024warm,lovell2024halo,yung2024ultra}. We refer interested readers to \citet{morales2010reionization} and \citet{dayal2018early} for recent reviews. This work focuses on the end stages of reionization when the IGM has been predominately heated and ionized.

\RefereeReport{Several previous models of reionization, such as \citet{dixon2016large}, \citet{semelin201721ssd}, \citet{ross2019evaluating}, \citet{eide2020large} and \citet{ocvirk2020cosmic}, ended at $z\approx 6$ as the observational constraints \citep[e.g.][]{fan2006constraining} showed very small neutral fractions ($\lesssim 10^{-3}$) below this redshift.
Recent studies have updated our understanding of the end stages of reionization, hinting that the residual neutral islands can have significant impact on the quasar spectra observations during $z\lesssim 6$ \citep[e.g.,][]{kulkarni2019large,keating2020long,nasir2020observing}.  In subsequent works, the observations have focused on constraining the reionization history at these very late stages of reionization \citep[e.g.][]{bosman2021comparison,bosman2022hydrogen,zhu2021chasing,fan2023quasars}}. Observations have also suggested a substantial evolution in global quantities, such as the mean ionizing (UV) photon background \citep[e.g.,][]{wyithe2011near,calverley2011measurements} and the mean free path (MFP) of ionizing photons \citep[e.g.,][]{worseck2014giant,becker2021mean,bosman2021comparison,gaikwad2023measuring,zhu2023probing}, which has been challenging to reproduce in simulations. Recent studies have developed models focused on understanding these quantities \citep[e.g.,][]{d2018large,cain2021short,puchwein2019consistent,chardin2016large}. The models generated in this work are consistent with these latest measurements during the late stages \RefereeReport{of reionization}.

Hydrogen, the most abundant element in our Universe, can be probed by using the neutral hydrogen (\hi{}) 21-cm line emission, which can be measured with radio telescopes \citep[e.g.,][]{pritchard201221}. During the EoR, this signal can probe the evolution of the IGM, which depends on cosmological structure formation and the astrophysical processes of galaxy formation \citep[e.g.][]{iliev2012can,dixon2016large,Schaeffer2023beorn,schneider2023cosmological}. Previous studies have demonstrated its capability to constrain not only the properties of these early galaxies \citep[e.g.,][]{greig201521cmmc,mirocha2019does,park2019inferring,qin2021tale} but also aspects of cosmology \citep[e.g.,][]{sitwell2014imprint,kern2017emulating,Schneider2018constraining,liu2018implications,nebrin2019fuzzy,giri2022imprints}.

In the post-reionization era, most \hi{} resides inside galaxies, shielded from UV photons \citep[e.g.,][]{villaescusa2014modeling,villaescusa2018ingredients}. Unlike during the EoR, the 21-cm signal during this phase traces the large-scale distribution of galaxies. Therefore, this signal serves as a tracer of the large-scale structure of our Universe and has been proposed as a new cosmological probe \citep[e.g.][]{camera2013cosmology,xu2016precise,carucci2017imprints,wu2022prospects}. However, during the end stages of reionization, the 21-cm signal traces both the \hi{} in galaxies and the IGM \citep[e.g.][]{wyithe2008fluctuations,xu2019h}. Previous studies have explored this using semi-analytical frameworks and found that the residual \hi{} in the IGM can impact the signal at high-density regions \citep{miralda2000reionization,wyithe2008fluctuations}. These studies inferred an `outside-in' nature of reionization during the end stages. This suggests the ionization fraction is anti-correlated with matter density due to higher recombination rates in dense regions.
We will test this hypothesis using fully numerical simulations.


Ongoing radio experiments, such as the Low-frequency Array \citep[LOFAR;][]{mertens2020improved}, Murchison Wide-field Array \citep[MWA;][]{trott2020deep} and Hydrogen Reionization Array \citep[HERA;][]{hera2022upper,hera2023improved}, are improving the upper limits on the power spectrum of the 21-cm signal. These observations have provided valuable constraints on the early Universe physics \citep[e.g.][]{ghara2020constraining, ghara2021constraining, mondal2020tight, greig2021interpreting, greig2021exploring, hera2022constraints}. Among these measurements, \citet{trott2020deep} went down to $z\approx 6.5$, closest to the end stages of reionization. Additionally, the 21-cm signal has been detected during the post-reionization era in cross-correlation \citep{chime2023detection}. Cross-correlating the post-reionization 21-cm signal with other tracers of cosmological matter distribution can further constrain cosmology \citep[e.g.,][]{padmanabhan2020cross}.

In the near future, we expect to detect the post reionization 21-cm power spectrum with advanced radio experiments, including Baryon Acoustic Oscillations In Neutral Gas Observations \citep[BINGO;][]{abdalla2022bingo}, Five-hundred-meter Aperture Spherical radio Telescope \citep[FAST;][]{bigot2015hi}, and Hydrogen Intensity and Real-time Analysis eXperiment \citep[HIRAX;][]{newburgh2016hirax}.
The Square Kilometre Array (SKA) is currently under construction and will be powerful enough to go beyond the power spectrum, providing images of the \hi{} distribution \citep{mellema2015hi}. SKA will consist of two frequency components, SKA-Mid and SKA-Low, focusing on observing post-reionization and the EoR, respectively.
The initial phase of SKA-Low will cover a frequency range from about 50 to 350 MHz\footnote{The latest details about the SKA can be found at \url{https://www.skao.int/en/explore/telescopes}.}, corresponding to redshifts between 30 and 3 \citep{koopmans2015cosmic}. Thus, this component will be instrumental in probing the end stages of reionization. The suite of simulations developed in this work will play a vital role in understanding these measurements. 

Modelling the IGM during these end stages is challenging because almost all the ionized bubbles are overlapping, making large distances in the IGM transparent to UV photons \citep[e.g.][]{xu2017islandfast,giri2019neutral}. Additionally, we need to model the astrophysical processes inside galaxies to study the amount of \hi{} shielded from UV photons. In principle, a hydro-dynamical N-body framework with radiative transfer calculation would be capable of fully depicting the transition from EoR to Post-EoR. However, such simulations are computationally challenging due to the large dynamical range required to study the \hi{} in the IGM and inside galaxies. Computationally cheap semi-numerical frameworks become inaccurate during these times as bubble overlaps lead to photon non-conservation \citep[e.g.][]{choudhury2018photon}. In \citet{giri2019neutral}, we developed a suite of large-scale radiative transfer simulations to study the IGM during these end stages. We will extend this exploration with a new simulation suite tuned to the latest measurements and including the post-reionization \hi{} distribution.

This paper is structured as follows: in the next section, we describe our simulation framework. We present our models in Section~\ref{sec:models} and discuss the simulated 21-cm signal in Section~\ref{sec:21cm}. Finally, we summarise our findings in Section~\ref{sec:conclusion}.

\section{Simulation framework} \label{sec:methods}

We first describe the \nbody{} simulation used in this study to model the photon sources during the epoch of reionization. 
To explore the evolution of the 21-cm signal during the end stages of reionization, we require modelling of both the \hi{} distribution in the IGM and the \hi{} inside galaxies that are self-shielded from UV photons.
Therefore, in subsection~\ref{sec:reion_model}, we define the sources of reionization and the method to model unresolved sinks in the IGM. Finally, in subsection~\ref{sec:HI_inside_haloes}, we explain our method for modelling the \hi{} content inside galaxies.

\subsection{Cosmological structure formation}
We model the cosmological structure formation by running an \nbody{} code, \pkdgrav{} 
\citep{potter2017pkdgrav3}, with $2048^3$ particles in a simulation box of length $298$ cMpc in each direction. We chose this box size to reduce the sample variance at large scales \citep[i.e. $k\approx 0.1\,\mathrm{Mpc}^{-1}$;][]{iliev2014simulating,giri2023suppressing}, where the current radio experiments provided the best upper limits. This simulation setup gives us a particle mass of $m_{\rm part} = 1.2 \times 10^8 M_\odot$. We identify dark matter haloes using a friends-of-friend halo finder implemented in \pkdgrav{} with a minimum of 10 dark matter particles, which helped resolve haloes down to the high-mass atomically cooling haloes (HMACHs) defined in \citet{dixon2016large}. Though this minimum number of particles is particularly low, we found a good match with the halo mass function calculated from extended Press-Schechter formalism. We show this validation test in Appendix~\ref{app:hmf}. We assume that star formation in haloes smaller than HMACHs is suppressed, especially during the end stages of reionization, due to radiative feedback \citep[e.g.][]{nebrin2023starbursts}. 

Throughout this study, we have considered a flat cold dark matter cosmological model with parameters aligned with the \texttt{Planck 2018} results \citep{collaboration2020planck}. These parameters include a matter abundance of $\Omega_{\rm m}=0.32$, a baryon abundance of $\Omega_{\rm b}=0.044$, a dimensionless Hubble constant of $h=0.67$, and a standard deviation of matter perturbations at the 8$h^{-1}$ cMpc scale denoted as $\sigma_{\rm 8}=0.83$. We consider a primordial gas with Helium abundance mass factor of $Y_p=0.249$, which gives a molecular weight of $\mu=(1-Y_p)^{-1}=1.33$. We initialised our \nbody{} simulation at $z=150$ with second-order Lagrangian perturbation theory \citep[e.g.][]{bertschinger1998simulations} using a transfer function generated with the Boltzmann code, \texttt{class} \citep{lesgourgues2011cosmic,blas2011cosmic}. The \nbody{} simulation was run down to $z\approx 4$ while saving snapshots at every $\sim$10.86 Myrs. We used the Piecewise Cubic Spline mass assignment scheme \citep{sefusatti2016accurate} to generate matter density fields on $256^3$ grids.

\subsection{Reionization of the intergalactic medium}\label{sec:reion_model}
We simulate the evolution of the IGM using our state-of-the-art radiative transfer simulation code \pyccray{} \citepalias{hirling2024pyc}, which is an upgraded version of \ccray{} \citep{mellema2006c2}. This code has a user-friendly \texttt{Python} interface and it can leverage Graphics processing units (GPUs) to reduce the computing time of solving the three-dimensional radiative transfer equation in a cosmological simulation volume. 

To leverage the latest measurements of the high-redshift sources, we have implemented a flexible parameterization designed to identify viable ionizing source models. We detail this parameterization in subsection~\ref{sec:reion_source_model}.
During late times, the ionization state of the IGM may depend on small-scale absorbers.
In subsection~\ref{sec:reion_sink_model}, we describe how we deal with the absorption of photons by unresolved density fluctuations.

\subsubsection{Source model}\label{sec:reion_source_model} 
We populate the dark matter haloes provided by our \nbody{} simulation with galaxies of stellar mass $M_\star=f_\star(M_h) (\Omega_\mathrm{b}/\Omega_\mathrm{m}) M_h$, where $f_\star$ describes the fraction of baryonic mass falling into a halo of mass $M_h$ that gets converted to stars. 
We parameterize this star formation inside the haloes using the following \textit{stellar-to-halo} relation,
\begin{eqnarray}
    f_\star(M_\mathrm{h}) \equiv \frac{M_\star}{(\Omega_\mathrm{b}/\Omega_\mathrm{m})M_\mathrm{h}} = f_{\star,0} \left(\frac{M_\mathrm{h}}{10^{10}M_\odot}\right)^{\alpha_\star}
    \ ,
    \label{eq:f_star_reduced}
\end{eqnarray}
where $f_{\star,0}$ and $\alpha_\star$ are the normalization constant and index of the power-law, respectively. We should note that our simulations framework, \pyccray{}, implements a more generic parametrization provided in \citet{schneider2021halo,schneider2023cosmological}, which is needed for studying non-cold dark matter models \citep{giri2022imprints}. In the current work, we consider the reduced form given above that is enough to model the ultraviolet luminosity functions \citep[UVLFs; e.g.][]{gillet2020combining,park2020properties}.

We require the photons produced inside the sources to model the gas properties in the IGM. We model the rate of the number of UV photons ($\dot{N}_\gamma$) escaping into the IGM with the following relation, 
\begin{eqnarray}
\dot{N}_\gamma(M_h) = f_\mathrm{esc}(M_h) \left(\frac{N_\mathrm{ion}}{m_\mathrm{p}}\right)\dot{M}_\star(M_h)
\,
\label{eq:N_gamma}
\end{eqnarray}
where $f_\mathrm{esc}$, $N_\mathrm{ion}$ and $\dot{M}_\star$ are the escape fraction, ionizing photon production efficiency and stellar mass growth, respectively. \RefereeReport{$m_\mathrm{p}$ is the proton mass and $N_\mathrm{ion}$ gives}
the number of ionizing photons produced inside a source per unit proton mass. 
\RefereeReport{We assume that} the mass dependence of the escape fraction can be parameterized with a power-law given as \citep[e.g.][]{park2019inferring},
\begin{eqnarray}
    f_\mathrm{esc} (M_h) = f_\mathrm{esc,0}\left(\frac{M_h}{10^{10}M_\odot}\right)^{\alpha_\mathrm{esc}} \ ,
\end{eqnarray}
where $f_\mathrm{esc,0}$ and $\alpha_\mathrm{esc}$ are free parameters.

We model the star formation rate as $\dot{M}_\star = f_\star (M_h) (\Omega_\mathrm{b}/\Omega_\mathrm{m}) \dot{M}_h$, where $\dot{M}_h$ is the halo accretion rate. We assume an \textit{exponential} mass growth model\footnote{A similar accretion model is implemented in the \texttt{21cmFAST} framework, but their timescales differ by a factor of $(1+z)$. See eq.~3 in \citet{park2019inferring} for comparison. See \citet{schneider2021halo} for a comparison of several mass growth models.} 
where the accretion rate is given as,
\begin{eqnarray}
    \dot{M}_h(z) = \frac{M_h(z)}{[\alpha_h (1+z) H(z)]^{-1}}  \ .
    \label{eq:Mh_dot}
\end{eqnarray}
$H(z)$ is the Hubble parameter and $\alpha_h$ is a free parameter. We set $\alpha_h=0.79$ to obtain a good match to detailed simulations from  \citet{behroozi2020universe}. This mass growth model incorporates both the gradual halo growth and mergers \citep[e.g.][]{mcbride2009mass,dekel2013toy,trac2015scorch}.

For modelling the reionization of the IGM, \pyccray{} reads the halo catalogue from the \nbody{} simulation and calculates the corresponding $\dot{N}_\gamma$ using Eq.~\ref{eq:N_gamma}. This quantity is then assigned to the relevant grid cells in the gridded matter density fields. The transfer of radiation in the IGM is modelled keeping $\dot{N}_\gamma$ fixed between successive time steps of the \nbody{} snapshots.

\subsubsection{Sink model}\label{sec:reion_sink_model} 
Large-scale simulation volumes such as ours cannot resolve the small-scale density structures relevant for absorbing UV photons. Several physical interpretations exist of these unresolved sinks, including self-shielded systems (i.e. Lyman-limit systems) and small-scale density fluctuations that can increase recombination. Modelling these structures is crucial for this work as the IGM in the late stages are sensitive to the properties of the sinks \citep[e.g.][]{cain2021short,cain2024rise,gaikwad2023measuring}. Previous authors exploring this problem have suggested different methods to model them \citep[e.g.][]{miralda2000reionization,ciardi2006effect,choudhury2009inside,sobacchi2014inhomogeneous,shukla2016effects,mao2020impact,bianco2021impact}. 

We have implemented multiple approaches to account for the effect of unresolved sinks in \pyccray{}, such as through the clumping factor $\mathcal{C}$ and mean free path $R_\mathrm{mfp}$. The former approach boosts the recombinations by increasing the ratio of variance of the matter density and the square of the mean density. The latter approach defines a maximum allowed distance for the photons to travel in the IGM, assuming these photons get absorbed by unresolved sinks. We should note that the ionized regions in the IGM can have sizes larger than $R_\mathrm{mfp}$ due to ionized bubble overlaps.

In our framework, we have implemented a redshift evolving $R_\mathrm{mfp}$, which is parameterised as,
\begin{eqnarray}
    R_\mathrm{mfp}(z) = A_\mathrm{mfp} \left(\frac{1+z}{5}\right)^{\eta_\mathrm{mfp,0}} \left[1+\left(\frac{1+z}{1+z_\mathrm{mfp}}\right)^{\eta_\mathrm{mfp,1}}\right]^{\eta_\mathrm{mfp,2}} \ ,
    \label{eq:R_mfp}
\end{eqnarray}
where $A_\mathrm{mfp}$, $z_\mathrm{mfp}$, $\eta_\mathrm{mfp,0}$, $\eta_\mathrm{mfp,1}$ and $\eta_\mathrm{mfp,2}$ are free parameters. The above expression is a modified form of the fit provided in \citet{worseck2014giant}, which can be reconstructed by setting $\eta_\mathrm{mfp,2}=0$. A non-evolving $R_\mathrm{mfp}$ can be modelled by setting both $\eta_\mathrm{mfp,1}$ and $\eta_\mathrm{mfp,2}$ to zero. We introduced this modification to incorporate recent measurements of $R_\mathrm{mfp}$, which will be presented in Section~\ref{sec:absorbers}.

\subsection{Neutral hydrogen inside haloes}\label{sec:HI_inside_haloes} 
After the reionization of the IGM, the 21-cm signal is produced by \hi{} that persists within massive galaxies, having self-shielded against the UV background.
Direct observation of this \hi{} in galaxies beyond $z\sim 0.4$ through 21-cm emission lines is quite challenging, and thus, observing the integrated 21-cm line of unresolved galaxies over a wide sky area can provide a more efficient and cost-effective probe for cosmological matter distribution \citep{Chang2010Natur,bagla2010h}. Conversely, simulating the \hi{} content in galaxies is extremely complex, requiring high-resolution hydro-dynamical N-body simulations coupled with radiative feedback and cooling mechanisms \citep{Bharadwaj2004III, Bird2014Damp}. However, a reasonable approximation of the averaged \hi{} content can be obtained based on the hosting halo mass, $M_\mathrm{h}$, and several models have been proposed in the literature \citep[see, e.g.,][]{villaescusa2014modeling,villaescusa2018ingredients,padmanabhan2017halo,Modi2019Int,Spinelli2020}.

Therefore, we adopt a straightforward approach and assign the mass of \hi{} ($M_\mathrm{HI}$) to the haloes in our simulation using the $M_\mathrm{HI}-M_\mathrm{h}$ relation from \citet{padmanabhan2017halo}, given as
\begin{eqnarray}
    M_{\text{HI}}(M_\mathrm{h}) = \alpha\, f_{H,c}\,M_\mathrm{h}\, \left(\frac{M_\mathrm{h}}{10^{11}h^{-1}M_\odot}\right)^\beta \exp\left[-\left(\frac{v_{c,0}}{v_c(M_\mathrm{h})}\right)^3\right] \ .
    \label{eq:MHI_Mh}
\end{eqnarray}
Here, $\alpha$ is a normalisation factor and represents the fraction of \hi{}, relative to cosmic hydrogen mass fraction, $f_\mathrm{H,c} = (1 - Y_p)\Omega_b / \Omega_m$, associated with a dark matter halo of mass $M_\mathrm{h}$. $\beta$ is the logarithmic slope of the $M_\mathrm{HI}-M_\mathrm{h}$ correlation. $v_{c,0}$ is the virial velocity below which the neutral hydrogen is supposed to be suppressed. We choose $\alpha = 0.9$, $\beta = -0.58$ and $v_{c,0} = 10^{1.56}\,\rm km/s$, as used by \cite{padmanabhan2017halo}, which was based on constraints from observational data of the abundance and clustering of \hi{} in galaxies between redshift $z\sim 0$ and $5$. We calculate the virial velocity of the halos, $v_c(M)$, based on the relation:
\begin{equation}
    v_c(M_\mathrm{h}) = 96.6\, {\rm km\,s^{-1}}\, \left(\frac{\Delta_c\,\Omega_m}{54.4}\right)^{\frac{1}{6}}\, \left(\frac{1+z}{3.3}\right)^{\frac{1}{2}} \left( \frac{ M_\mathrm{h} }{10^{11}\,M_\odot}\right)^{\frac{1}{3}}
\end{equation}
where $\Delta_c = 178$ is the mean over-density of the halo \citep{Maller2004, Barnes2014DLA}. This paper will explore the evolution of the 21-cm signal using gridded simulation volumes conducted on a mesh with a low spatial resolution ($\Delta r \sim 1.17$ cMpc). Therefore, after assigning a value of $M_\mathrm{HI}$ to the haloes at different redshifts, we place them on the same grids of the matter density distribution used to model the reionization of the IGM.
\RefereeReport{We should note that this assignment process does not self-consistently incorporate the impact of the UV radiation background created during reionization on the \hi{} content inside haloes. The parameterisation in Equation~\ref{eq:MHI_Mh} models this feedback process, which has been compared to hydrodynamical simulations \citep[e.g.][]{villaescusa2018ingredients}.}

\section{Reionization models} \label{sec:models}
\begin{table*}
    \centering
    \caption{A list of reionization models along with the corresponding parameters. These parameters are grouped into source and sink parameters. We also mention the Thompson optical depth ($\tau_\mathrm{e}$) for our models that are consistent with the \texttt{Planck 2018} constraints \citep{collaboration2020planck}.}
    \begin{tabular}{cccccccccccc}
        \toprule
        Model Name & \multicolumn{3}{c}{Source} & \multicolumn{5}{c}{Sink} & Thompson optical depth\\
        \cmidrule(lr){2-4} \cmidrule(lr){5-9} \cmidrule(lr){10-10}
         & $N_\mathrm{ion}\, f_{\star,0}\, f_\mathrm{esc,0}$ & $\alpha_\star$ & $\alpha_\mathrm{esc}$ & $A_\mathrm{mfp}$ & $z_\mathrm{mfp}$ & $\eta_\mathrm{mfp,0}$ & $\eta_\mathrm{mfp,1}$ & $\eta_\mathrm{mfp,2}$ & $\tau_\mathrm{e}$\\
        \midrule
        Source1\_SinkA & 4 & 0.3 & 0.0 & 210 & 6.0 & -9 & 9 & 1 & 0.049 \\
        Source2\_SinkA & 4 & 0.3 & -0.3 & 210 & 6.0  & -9 & 9 & 1 & 0.051 \\
        Source3\_SinkA & 4 & 0.3 & -0.5 & 210 & 6.0  & -9 & 9 & 1 & 0.053\\
        Source1\_SinkB & 4 & 0.3 & 0.0 & 210 & 5.5  & -9 & 9 & 1 & 0.049\\
        \bottomrule
    \end{tabular}
    \label{tab:models}
\end{table*}

\begin{figure*}
    \centering
    \includegraphics[width=1.0\linewidth]{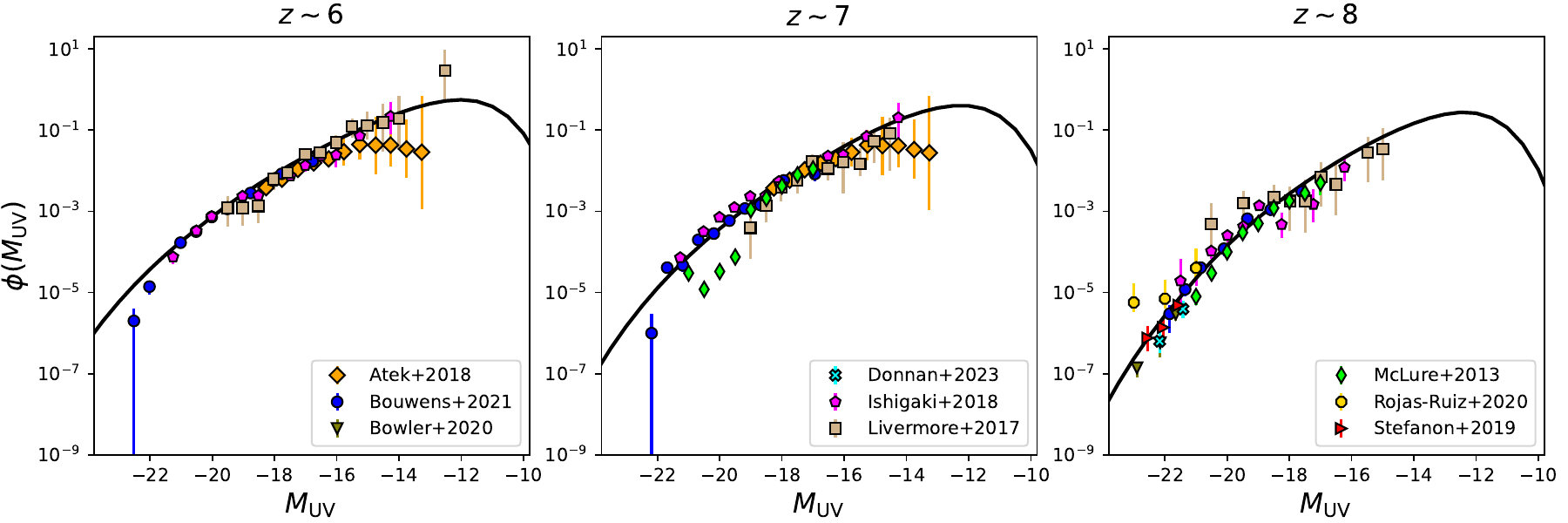}
    \caption{The ultraviolet luminosity function (UVLF) gives the number density of photon sources of different brightness ($M_\mathrm{UV}$). We plot several measurements \citep{atek2018extreme,bouwens2021new,bowler2020lack,donnan2023evolution,ishigaki2018full,livermore2017directly,mclure2013new,rojas2020probing,stefanon2019brightest} binned at three different redshifts ($z=6,7$ and 8). The solid line represents the best-fit model of the \textit{stellar-to-halo} relation (Eq.~\ref{eq:f_star_reduced}) against these measurements.
    }
    \label{fig:source_uvlf}
\end{figure*}

\RefereeReport{The reionization of the IGM is not well constrained, as it has not been directly observed. Therefore,}
we develop multiple models with properties consistent with available measurements. First, we describe the source and sink parameter values of our cosmic reionization models in subsection~\ref{sec:model_params}. In subsection~\ref{sec:reion_hist}, we describe the reionization history of the IGM in our models. Later, in subsection~\ref{sec:topology} and~\ref{sec:uv_background}, we study the evolution of the topology of neutral islands distribution in the IGM and the UV background, respectively, in these models. 

\subsection{Model parameters}\label{sec:model_params}
Here we will use the latest observations to find relevant UV photon source and sink model parameters.

\subsubsection{Stellar mass} \label{sec:stellar_mass}
We begin by finding suitable intrinsic source properties for our simulations by using the ultraviolet luminosity function (UVLF) measurements, which give the number density of sources at various UV brightness. We follow \citet{park2019inferring} to express this quantity as
\begin{eqnarray}
  \phi_{\rm UV} = f_\mathrm{duty} (M_h)\frac{\mathrm{d} n}{\mathrm{d} M_{h}}\frac{\mathrm{d} M_{h}}{\mathrm{d} M_{\rm UV}} \ ,
  \label{eq:UV_luminosity}
\end{eqnarray}
where $({\mathrm{d} n}/{\mathrm{d} M_{h}})$ and $M_\mathrm{UV}$ are the halo mass function (HMF) and the absolute UV magnitude, respectively. $f_\mathrm{duty}=\exp (-M_\mathrm{turn}/M_h)$ is the duty cycle that models
\RefereeReport{the absence of haloes below a mass of $M_\mathrm{turn}$. We should note that this parameter degenerate with  to the suppression of star formation in small mass haloes, which can occur due to radiative feedback \citep[e.g.][]{nebrin2023starbursts}.}
$M_\mathrm{UV}$ quantifies the brightness of sources that are given as
\begin{multline}
  M_\mathrm{UV} = M_0 - 2.5 \biggr[ \log_\mathrm{10}f_\star(M_h,z) + \log_\mathrm{10}\frac{\Omega_b}{\Omega_m} \\
  + \log_\mathrm{10}\frac{\dot{M}_h(M_h,z)}{\rm M_\odot yr^{-1}} - \log_\mathrm{10}\frac{\kappa}{\rm M_\odot yr^{-1}} \biggr] \ ,
  \label{eq:Muv}
\end{multline}
where $M_0 = 51.6$ corresponding to AB magnitude system \citep{Oke1974ABsystem}. We use $\kappa = 1.15 \times 10^{-28} ~{\rm M_\odot yr^{-1}/(erg~ s^{-1} Hz^{-1}})$, calibrated for 1500 \AA~dust-corrected rest-frame UV luminosity. This calibration assumes continuous star formation and a Salpeter stellar initial mass function \citep{madau2014cosmic}.

Figure~\ref{fig:source_uvlf} shows measurements of UVLFs that are binned at three redshifts ($z=6$, 7 and 8). Using values of $f_{\star,0}=0.1$ and $\alpha_\star=0.3$, we find good fits to these measurements and adopt these values for our source model. The truncation at the faint end is caused by the duty cycle with $M_\mathrm{turn}=10^9 M_\odot$, which is set by the smallest halo mass in our simulations. \RefereeReport{The figure reveals that these UVLF measurements are not sensitive enough to explore luminous sources residing in haloes below $10^9 M_\odot$.}
We should note that $f_{\star,0}$ degenerates with $N_\mathrm{ion}$, which is currently loosely constrained by measurements. The parameter $\alpha_\star$ models the mass-dependence of our source model and allows the construction of star formation models in which either low-mass or high-mass halos dominate the process. In all our source models, we keep the parameters of the $f_\star(M_h)$ relation fixed as the UV photon budget will include the escape fraction model, which will be discussed next.

\subsubsection{ionizing photon budget} \label{sec:uv_budget}
\begin{figure}
    \centering
    \includegraphics[width=0.95\linewidth]{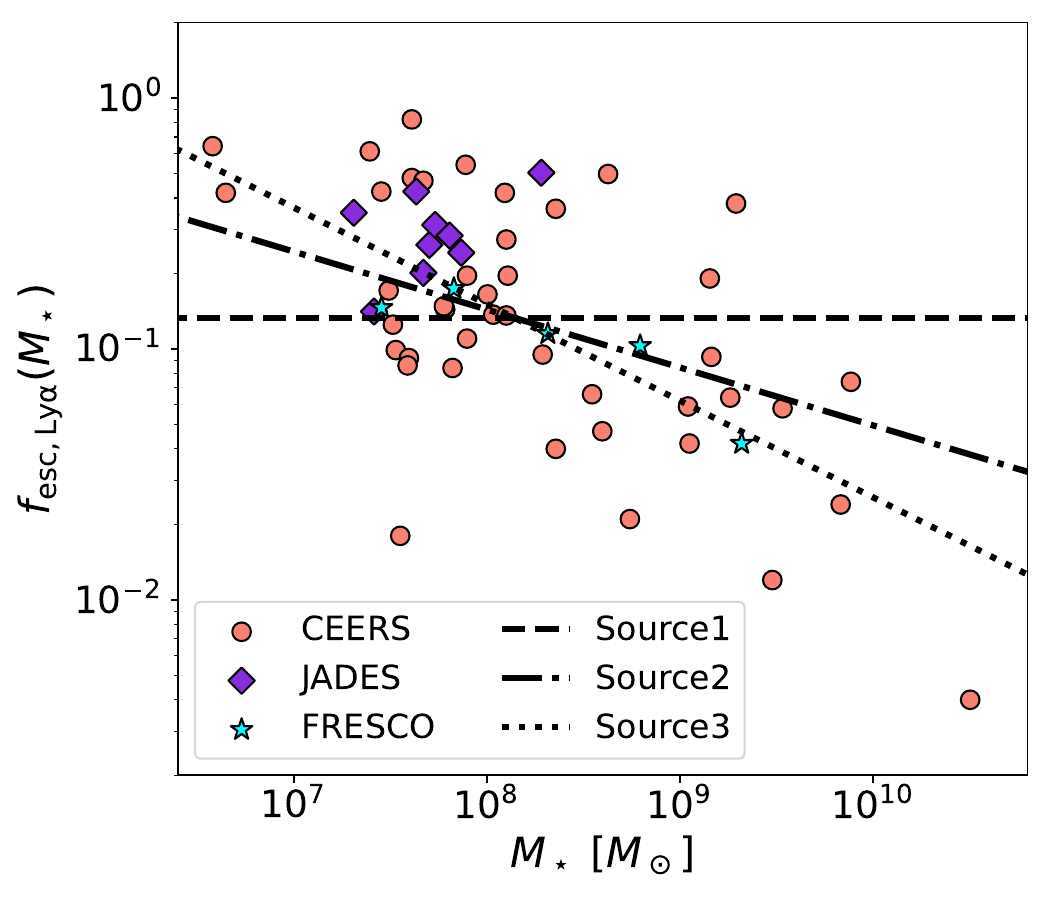}
    \caption{The \lya{} escape fraction plotted against various stellar masses. We plot three models (black lines) with different mass dependence of the escape fraction that are consistent with estimations from CEERS (circles), JADES (diamonds) and FRESCO (stars) survey data \citep[][]{lin2024quantifying,Napolitano2024Lya}. 
    We should note that these data are provided at different redshifts between $4.5\lesssim z \lesssim 8.5$.
    We estimate the UV escape fraction through the relation $f_\mathrm{esc}\approx 0.15 f_\mathrm{esc,Ly\alpha}$ \citep{begley2024connecting}.}
    \label{fig:f_esc}
\end{figure}

In our framework, the total rate of escaping UV photons from a source is defined by Eq.~\ref{eq:N_gamma}. While the star formation rate is well constrained by observations of the UVLF (discussed in the previous subsection), the photon escape rate is much harder to constrain. We have some indirect constraints on the escape fraction of UV photons through the \lya{} photon escape, given the strong correlation between \lya{} and UV escape \citep{chisholm2018accurately}. For the comparison here, we assumed a relation for UV escape fraction $f_\mathrm{esc}\approx 0.15 f_\mathrm{esc,Ly\alpha}$ \citep{begley2024connecting}, where $f_\mathrm{esc,Ly\alpha}$ is the \lya{} escape fraction.

The \lya{} photon escape can be estimated using the features in the high redshift spectroscopic data, such as the UV slope and line strengths \citep[e.g.][]{zackrisson2013spectral,zackrisson2017spectral,jensen2016machine,jaskot2019new,giri2020identifying,begley2022vandels}. 
In Figure~\ref{fig:f_esc}, we present some of the latest constraints on $f_\mathrm{esc,Ly\alpha}$ at several stellar masses $M_\star$. Though these measurements have a wide redshift range ($4.5\lesssim z \lesssim 8.5$), we consider them together. With the $M_\star(M_h)$ relation fixed against the UVLFs, we determine three models for the UV escape fraction, $f_\mathrm{esc}(M_h)$. We first set $f_\mathrm{esc,0}$ to 0.02, and consider models with $\alpha_\mathrm{esc}$ varied as 0 (Source1), -0.3 (Source2) and -0.5 (Source3). These models yield three different mass dependencies for the UV escape fraction. We list the above three distinct source models as the first three entries in Table~\ref{tab:models}. 

In all cases, the intrinsic ionizing efficiencies of the sources denoted as $N_\mathrm{ion}$, are kept fixed at 2000, which gives $N_\mathrm{ion}f_\mathrm{\star,0} f_\mathrm{esc,0} = 4$. 
We made this choice to obtain a consistent reionization history, which will be shown in Section~\ref{sec:reion_hist}. Notably, the three UV source models, $\dot{N}_\gamma (M_h)$, exhibit significantly different mass dependencies. 
The final mass dependence of $\dot{N}_\gamma$ is given by $\alpha_\star+\alpha_\mathrm{esc}$.
While the source model in `Source2' is independent of halo mass, the $\dot{N}_\gamma$ in `Source1' and `Source3' are correlated and anti-correlated, respectively, with halo masses. We will study the impact of these mass dependencies in the next section. The fourth model listed in Table~\ref{tab:models} has the same source property as `Source1' but with a different sink model. We will discuss the different sink models in the following subsection.

\subsubsection{Unresolved absorbers} \label{sec:absorbers}

\begin{figure}
    \centering
    \includegraphics[width=0.95\linewidth]{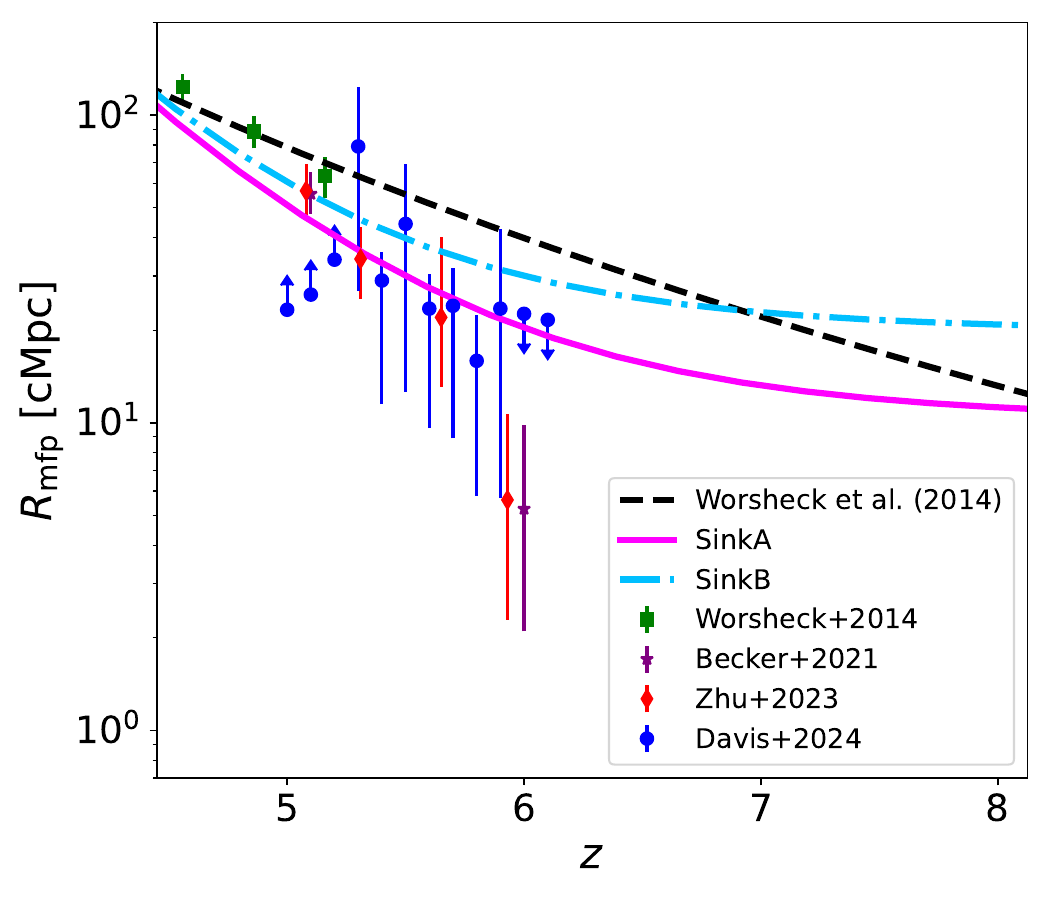}
    \caption{The mean free path models (depicted as lines) compared against various measurements \citep{worseck2014giant,becker2021mean,zhu2023probing,davies2023constraints}. The fit provided in \citet{worseck2014giant}, shown with a dashed line, does not agree with recent constraints. 
    The `SinkA' and `SinkB' models considered here using the modified parameterization (Eq.~\ref{eq:R_mfp}) have a better agreement with these constraints and converge to 10 and 20 cMpc at high redshifts.}
    \label{fig:sink_qso}
\end{figure}

In this study, we set $\mathcal{C}=1$ and explore the impact of sink models using the redshift dependent evolution of $R_\mathrm{mfp}$ defined in Eq.~\ref{eq:R_mfp}. Previous studies have used the fit from \citet{worseck2014giant} in their reionization simulations \citep[e.g.][]{shukla2016effects,choudhury2021studying}.
We can obtain this fit by setting $A_\mathrm{mfp}=175$ cMpc, $\eta_\mathrm{mfp,0}=-4.4$ and $\eta_\mathrm{mfp,2}=0$, which match the measurements at $z\approx 5$. However, Figure~\ref{fig:sink_qso} indicates that this fit overestimates the mean free path at $z\gtrsim 5$ compared to the latest constraints on $R_\mathrm{mfp}$. These data points, suggesting a stronger redshift dependence, lead us to adopt $A_\mathrm{mfp}=210$ cMpc and $\eta_\mathrm{mfp,0}=-9$ for a better match. 

It is important to note that with the power-law dependence of $R_\mathrm{mfp}$ provided in \citet{worseck2014giant} will approach zero at high redshifts ($z\gtrsim 7$), potentially hindering the formation of large ionized bubbles and violating reionization history constraints. We address this issue by introducing the modification shown in Eq.~\ref{eq:R_mfp}, which has three additional parameters, 
$z_\mathrm{mfp}$, $\eta_\mathrm{mfp,1}$ and $\eta_\mathrm{mfp,2}$. We set $\eta_\mathrm{mfp,1} = 9$ and $\eta_\mathrm{mfp,2} = 1$, exploring two models with $z_\mathrm{mfp} = 6$ and $z_\mathrm{mfp} = 5.5$. In these models, $R_\mathrm{mfp}$ converges to 10 and 20 cMpc, respectively, as evident in Figure~\ref{fig:sink_qso}. These models are marked as `SinkA' and `SinkB'. We have listed the sink parameter values in Table~\ref{tab:models}.

\subsection{Reionization history}\label{sec:reion_hist}
\begin{figure*}
    \centering
    \includegraphics[width=0.95\linewidth]{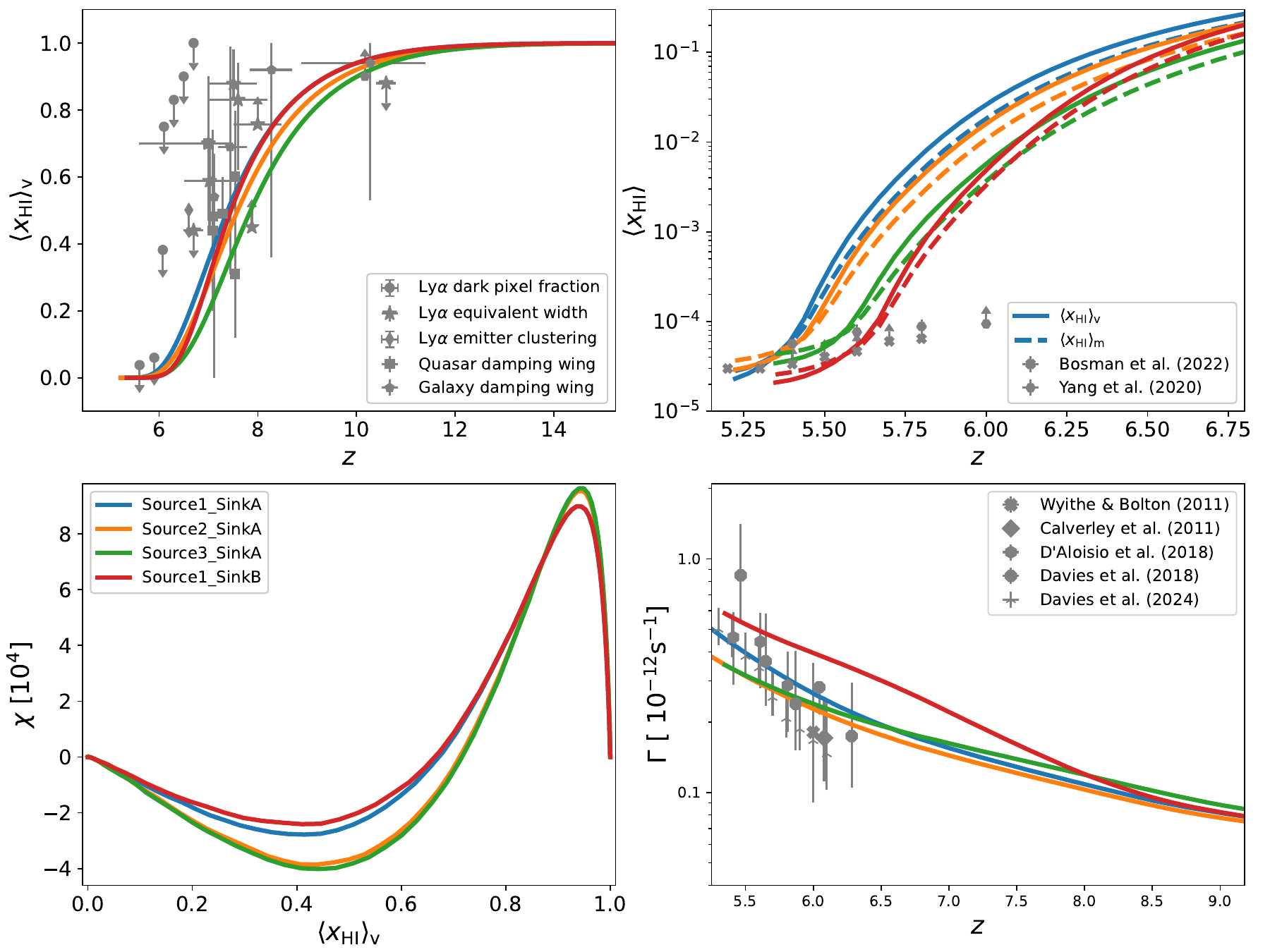}
    \caption{\textit{Top-left panel:} The intergalactic medium (IGM) reionization history, quantified by the volume-averaged neutral fraction \xHIv{}, of our models compared with constraints from several measurements. These measurements include \lya{} dark pixel fraction \citep[circles;][]{mcgreer2015model,jin2023nearly}, \lya{} equivalent width \citep[stars;][]{mason2018universe,mason2019inferences,hoag2019constraining,bolan2022inferring,bruton2023universe,morishita2023early,jones2023jades}, \lya{} emitter clustering \citep[diamonds;][]{ouchi2010statistics}, quasar damping wing \citep[squares;][]{davies2018quantitative,wang2020significantly,greig2022igm} and galaxy damping wing \citep[pentagons;][]{hsiao2023jwst,umeda2023jwst}. \textit{Top-right panel:} The volume-averaged (\xHIv{}; solid) and mass-averaged neutral fraction (\xHIm{}; dashed) at the end stages of reionization ($z\lesssim 7$). \RefereeReport{We show the measured constraints for these late stages derived from \lya{} forest data \citep{yang2020measurements,bosman2022hydrogen} for comparison.} Our models show a transition from \textit{inside-out} to \textit{outside-in} nature of reionization when \xHIv{}$\approx 5\times 10^{-5}$. 
    \textit{Bottom-left panel:} The plot shows Euler characteristics ($\chi$), which measures the topology of the distribution neutral islands in the IGM, evolving with \xHIv{}. All the reionization models have a distinct evolution, suggesting that the source and sink models affect the topology.
    \textit{Bottom-right panel:} The redshift evolution of the ionizing (UV) background compared against several constraints from quasar observations \citep{wyithe2011near,calverley2011measurements,d2018large,davies2018new,davies2023constraints}. This background grows over time until $z\approx 7$. Post this epoch, the value is governed by the choice of $R_\mathrm{mfp}$, which can be inferred from the slower growth at low redshifts. The model with `SinkB' has a background more significant than the models with `SinkA' due to larger $R_\mathrm{mfp}$ values at each redshift (see Figure~\ref{fig:sink_qso}).}
    \label{fig:reion_hist}
\end{figure*}

First, we investigate the reionization of the IGM in the different models simulated in this work. We show the volume-averaged neutral fraction (\xHIv{}) of the IGM in the top-left panel of Figure~\ref{fig:reion_hist}. Our models are carefully calibrated with the choice of $N_\mathrm{ion}$ to be consistent with a range of observational constraints on the reionization history. These constraints are represented with scattered points in the panel. The end of reionization is delayed in our models and falls well below $z\approx 6$, which is in line with the recent findings \citep[e.g.][]{kulkarni2019large,bosman2022hydrogen}.

The progress of reionization depends on our choice of source model. 
In the case of the `Source1' model ($\alpha_\star+\alpha_\mathrm{esc}>0$), the process is primarily driven by large-mass haloes. Conversely, in the `Source3' model ($\alpha_\star+\alpha_\mathrm{esc}<0$), small-mass haloes play a dominant role in driving reionization. The reionization history of models using `Source1' shows a delay compared to those using `Source3' due to 
\RefereeReport{brighter} 
small-mass haloes in the latter. The model employing `Source2' with $\alpha_\star+\alpha_\mathrm{esc}=0$ exhibits a reionization history that falls in between the previous two scenarios.

To understand the impact of the mean free path, we conducted simulations called `Source1\_SinkA' and `Source1\_SinkB,' which share the same source model but differ in the mean free path values. Initially, the reionization histories of these two models are identical. 
The $R_\mathrm{max}$ value in `SinkA' model is smaller than that in `SinkB' model at all times (see Figure~\ref{fig:sink_qso}).
However, in the middle stages, the reionization process accelerates in `Source1\_SinkB' compared to the other model, which can be understood from the larger value for $R_\mathrm{max}$. 
\RefereeReport{Thus, the unresolved sinks, modeled by defining a mean free path for ionizing photons, have} 
a substantial impact on the later stages of reionization.

In the top-right panel of Figure~\ref{fig:reion_hist}, we focus on the end stages of reionization ($z\lesssim 6.8$), plotting both volume-averaged (\xHIv{}) along with the mass-averaged (\xHIm{}) neutral fraction \RefereeReport{of the IGM}. Throughout most of the reionization process, we observe \xHIm{}$\lesssim$\xHIv{}, indicative of the \textit{inside-out} nature of reionization \citep{iliev2006simulating}. However, as reionization nears completion (\xHIv{}$\approx 5\times 10^{-5}$), we observe a transition to an \textit{outside-in} reionization pattern. During these very late stages, high recombination rates in dense regions dominate the reionization process, a phenomenon assumed in several previous studies \citep[e.g.,][]{miralda2000reionization,wyithe2008fluctuations}.

Additionally, we estimated the Thompson scattering optical depth ($\tau_\mathrm{e}$) derived from our models, which serves as a crucial constraint of the reionization history.
We have listed the values in Table~\ref{tab:models}. The \texttt{Planck} mission has precisely constrained this parameter to $0.054\pm 0.007$ at the 68 per cent confidence level \citep{collaboration2020planck}, representing a reduction compared to previous \texttt{WMAP} results \citep{hinshaw2013wmap}. This decline suggests a potential delay in the end of reionization \citep[e.g.][]{mitra2015cosmic}. Our models agree with these latest $\tau_\mathrm{e}$ constraints.

\subsection{Topology of neutral islands}\label{sec:topology}
To study the topology of the neutral island distribution in the IGM, we estimated the Euler characteristics ($\chi$) of our simulations, which describes the evolution of the connectivity of these islands due to the reionization process. 
We refer interested readers to \citet{friedrich2011topology} and \citet{giri2021measuring} for more discussion about topological evolution during reionization. 
The Euler characteristics is defined as
\begin{eqnarray}
    \chi = N_{\rm bubbles} - N_{\rm tunnels} + N_{\rm islands} \ ,
\end{eqnarray}
where $N_{\rm bubbles}$, $N_{\rm tunnels}$ and $N_{\rm islands}$ are the number of ionised bubbles, tunnels and neutral islands, respectively. 
\RefereeReport{We identify these structures in our simulations by applying a threshold of 0.5 to the ionization fractions in each cell.}
\RefereeReportTwo{During reionization, ionised bubbles connect to form complex structures \citep[e.g.][]{furlanetto2016reionization,giri2018bubble}. Tunnels of neutral regions develop within these complex bubbles, while islands are isolated neutral regions surrounded by ionised regions. Instead of directly counting the number of bubbles, tunnels and islands, we estimate $\chi$ by constructing a cubical complex, which triangulates these structures. This involves taking the alternating sum of the number of vertices, edges and faces \citep[e.g.][]{edelsbrunner2022computational}.}

In the bottom-left panel of Figure~\ref{fig:reion_hist} , we present $\chi$ for all our models at different epochs of reionization. 
At early times, $\chi$ increases due to the increasing number of ionized bubbles in the IGM. When small mass haloes drive reionization, a larger number of these ionized bubbles form. Therefore, `Source3' has the most prominent peak value of $\chi$. The sink models do not affect the topology at these early stages, as revealed by the overlapping evolution of the two sink models with the same source property.
Thus, we infer that the $\chi$ at early stages of reionization is useful in distinguishing between source models.

Over time, the ionized bubbles merge, creating tunnels that decrease $\chi$ and lead to a negative minimum during the intermediate stages. We observe differences in the behaviour of the two sink models during these stages. In the case of the `SinkB' model, which allows larger ionized bubbles to form for each source, this merger happens more quickly compared to the `SinkA' models. When connections occur with fewer number of bubbles, we get less number of tunnels, which is the case for `SinkB'. 
Consequently, the `Source1\_SinkB' model exhibits a minimum with the highest value.
This distinctive evolution of $\chi$ can help distinguish between different sink models.

In the late stages of reionization, $\chi$ depended on the number of neutral islands. Due to the small number of these islands, the magnitude of $\chi$ remains low during this period. We observed that the topology of the neutral islands tends to converge \RefereeReport{to similar values at $\langle x_\mathrm{HI}\rangle_\mathrm{v} \lesssim 0.1$} across all models at late times. 
\RefereeReport{We can measure this topology from the image data expected from SKA-Low with observation times as short as 100 hours \citep{giri2019neutral, giri2021measuring}. For studies on the impact of line-of-sight effects on the 21-cm signal image data, such as the light-cone effect and redshift-space distortion, we refer interested readers to \citet{giri2018bubble} and \citet{giri2021measuring}.}
Although the numerical count of these neutral islands is relatively small, their sizes remain notably significant, which will be further explored in Section~\ref{sec:21cm}.

\subsection{Ionizing photon background} \label{sec:uv_background}
We now study the mean ionization rate ($\Gamma$) within the \RefereeReport{ionized} IGM provided by our simulations. 
The bottom-right panel in Figure~\ref{fig:reion_hist} shows our results compared to observational constraints. Notably, these constraints predominantly apply to lower redshifts ($z\lesssim 6.5$). Replicating these constraints has posed a significant challenge in the field, often requiring abrupt changes in source properties towards the end stages of reionization \citep[e.g.,][]{chardin2016large,puchwein2019consistent,keating2020long,cain2021short,gaikwad2023measuring}. 
\RefereeReport{Our simulations, however, demonstrate a smooth evolution of $\Gamma$, achieved through our model for unresolved sinks and a gradual evolution of $R_\mathrm{mfp}$ without the need for significant changes in source properties. Several authors \citep[e.g.,][]{sobacchi2014inhomogeneous,nasir2021hydrodynamic,gaikwad2023measuring,davies2023constraints,fan2024cosmic,georgiev2024forest} have studied the strong dependence of this background on ionizing photon sinks. 
}

For a particular source model, the growth of $\Gamma$ is sensitive to the expansion of ionized bubbles merging together, rendering the IGM transparent to UV photons. During the final stages of reionization, most parts of the IGM become transparent to UV photons emitted by sources throughout the simulation volume. Therefore, we can consider the evolution of $\Gamma \propto R_\mathrm{mfp}$ \citep[e.g.,][]{haardt2012radiative} during this phase. In previous models \citep[e.g.,][]{dixon2016large,giri2019neutral}, we assumed a non-evolving $R_\mathrm{mfp}$ that resulted in the flattening of $\Gamma$ during late times. In this study, however, we assumed an increasing $R_\mathrm{mfp}$ during late times (see Section~\ref{sec:reion_sink_model}), causing $\Gamma$ to increase as reionization proceeds during the final stages. Furthermore, the evolution of $\Gamma$ observed in `SinkA' models closely resembles that seen in several hydro-dynamical simulations, such as \texttt{Thesan} \citep{garaldi2022thesan} and \texttt{CODA-III} \citep{lewis2022short}.

\section{The 21-cm signal}\label{sec:21cm}
This section presents the evolution of the 21-cm signal in our simulations. 
In subection~\ref{sec:bright_temp}, we begin by describing the quantity measured by radio telescopes and study the large-scale distribution of \hi{} during the end stages of reionization. In subection~\ref{sec:results_islands}, we discuss the existence of large neutral islands still remaining during this stage. Lastly, we present the power spectrum of the 21-cm signal in the concluding subsection.

\begin{figure*}
    \centering
    \includegraphics[width=0.99\linewidth]{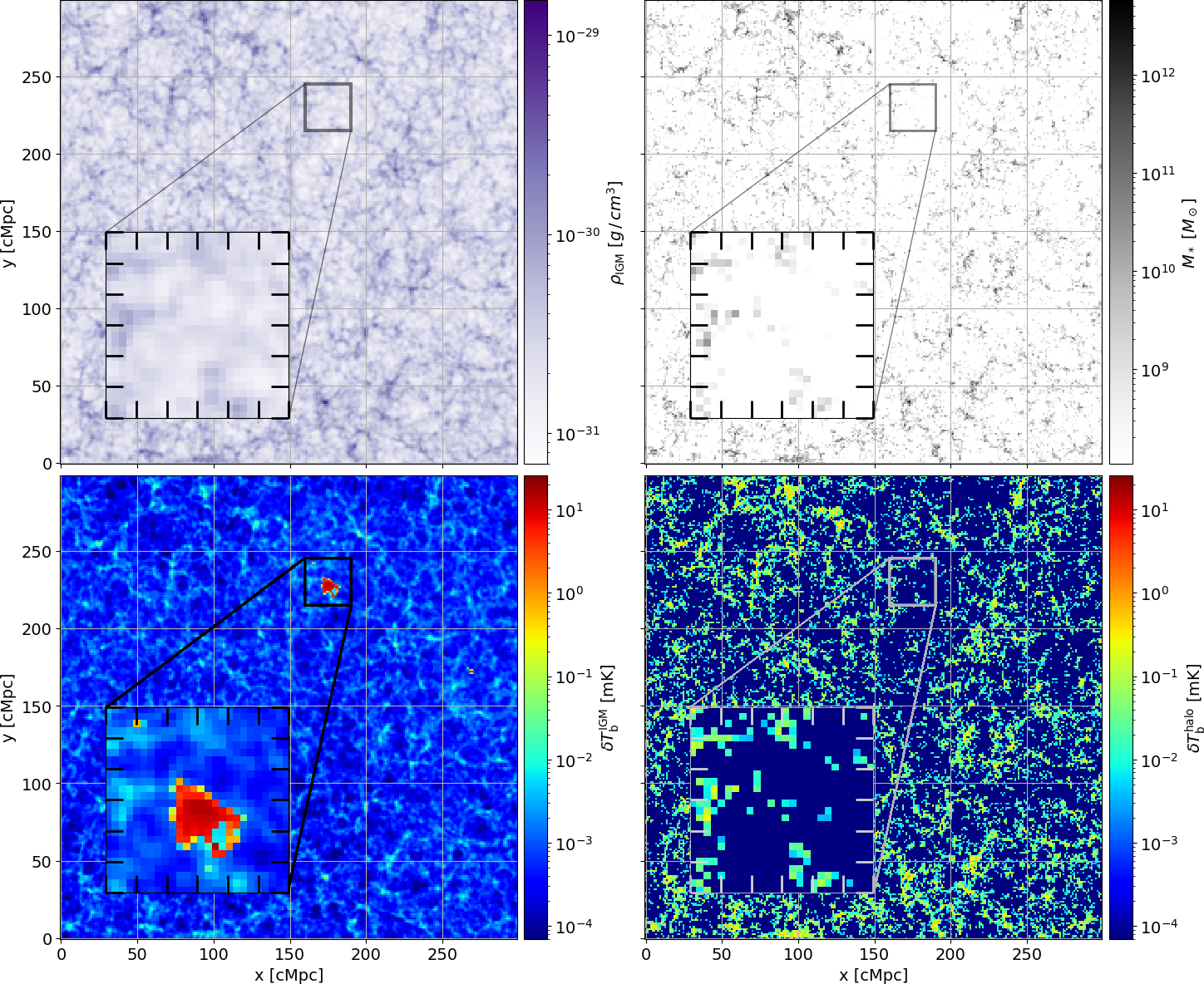}
    \caption{Slices from our model at redshift $z \approx 5.436$. 
    \textit{Top panels}: We show the snapshot from our \nbody{} run in volumes with a box length of 298 cMpc. The left and right panels depict the dark matter over-density ($\delta$) and cumulative stellar mass on the grid for our source model that was fixed using the UVLFs (see Section~\ref{sec:stellar_mass}) on the same grid, respectively. \textit{Bottom panels}: We illustrate the brightness temperature ($\delta T_\mathrm{b}$) corresponding to the 21-cm signal measured by radio telescopes. The left and right panels show this signal produced by the \hi{} in the intergalactic medium (IGM) of our Source1\_SinkA model and shielded \hi{} inside galaxies residing in the dark matter haloes. In all the panels, we zoom on to a region with a visibly large neutral island. Comparing these zoomed regions in the bottom panels, we infer that this neutral island is situated in a region with fewer haloes or the cosmic void.  
    }
    \label{fig:slices}
\end{figure*}

\subsection{Differential brightness temperature} \label{sec:bright_temp}

The 21-cm signal produced by \hi{} during reionization can be found at radio frequencies below 235 MHz.
The interferometry-based radio telescopes record the differential brightness temperature corresponding to this signal, which is given as \citep[e.g.][]{mellema2013reionization}
\begin{eqnarray}\label{eq:dTb}
\delta T_b(\mathbf{x},z)=T_0(z)x_{\rm HI}(\mathbf{x},z)\left[1+\delta_b(\mathbf{x},z)\right]
\left(1-\frac{T_{\rm cmb}(z)}{T_{\rm s}(\mathbf{x},z)}\right),
\end{eqnarray}
where $T_{\rm cmb}$ is the cosmic microwave background (CMB) temperature, and $x_{\rm HI}$ is the neutral hydrogen fraction. $\delta_b$ describes the \hi{} gas perturbation field that is assumed to follow the dark matter. The factor $T_0$ depends only on cosmology that is given by, 
\begin{eqnarray}
T_0(z) \approx 22 \left( \frac{\Omega_{\rm b} h^2}{0.022}\right) \left(\frac{0.144}{\Omega_{\rm m} h^2} \frac{1+z}{7}\right)^{1/2} {\rm mK}\ .
\label{eq:dT0}
\end{eqnarray}
$T_\mathrm{s}$ is the spin temperature of the gas, which we assume to be much higher than $T_\mathrm{CMB}$. 
Studies have shown that this spin temperature saturation occurs by the early stages of reionization \citep[e.g.][]{ghara201521,ross2017simulating,ross2021redshift}. Therefore it remains a reasonable assumption during the later stages of reionization and the post-reionization era.

\begin{figure}
    \centering
    \includegraphics[width=0.95\linewidth]{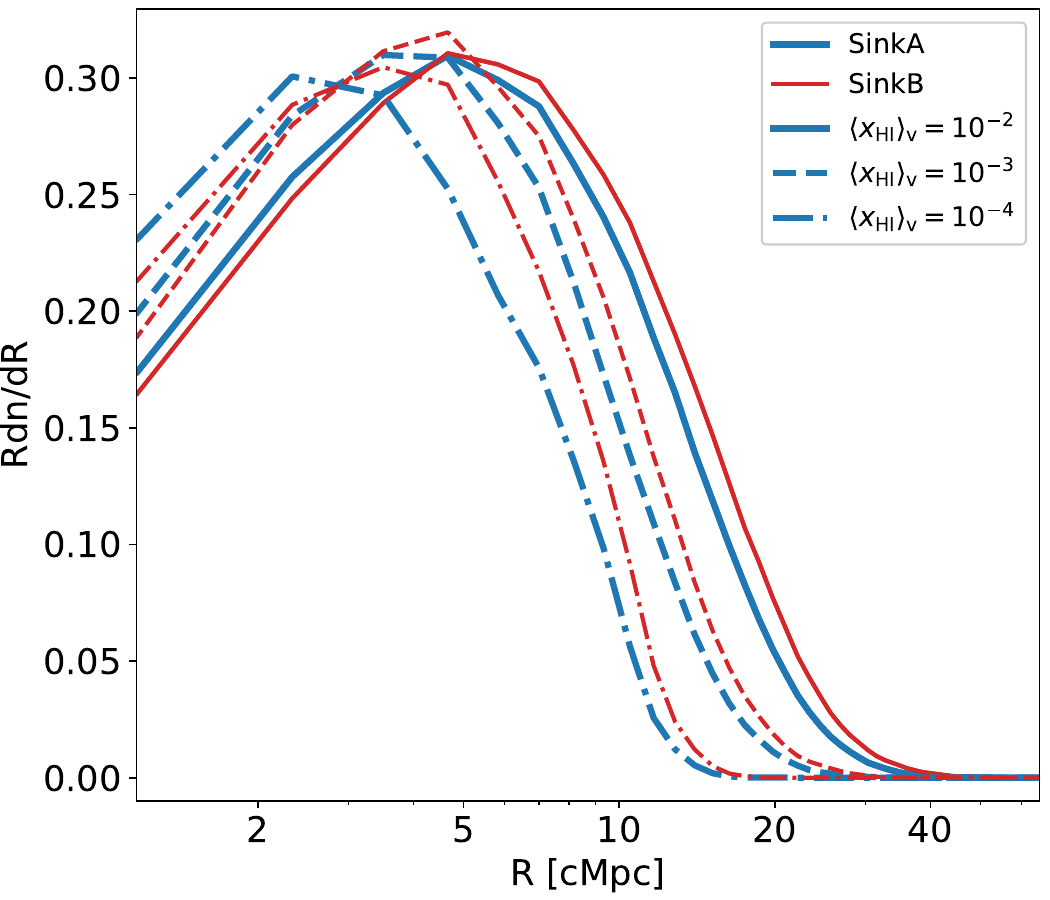}
    \caption{The size distribution of the neutral islands in `Source1\_SinkA' (blue) and `Source1\_SinkB' (red) models at three different volume-averaged neutral fractions \xHIv{}. These distributions show that significantly large neutral islands remain until the very late stages of reionization. 
    The distributions from `Source1\_SinkB' have the peaks positioned at larger sizes than `Source1\_SinkA' during each epoch.
    }
    \label{fig:ISDs}
\end{figure}

We first inspect the large-scale distribution of \hi{} in our simulation suite. Figure~\ref{fig:slices} illustrates slices from different fields at $z\approx 5.436$. The top-left panel shows the dark matter distribution. We assume the baryons to follow this distribution in IGM. In the top-right panel, we show the distribution of stellar masses. These gridded distributions are estimated by assigning stellar mass to dark matter haloes and summing them up for each grid point. The stellar masses were calculated using Eq.~\ref{eq:f_star_reduced} with the free parameters fixed against the UVLF observations (see Section~\ref{sec:stellar_mass}). We observe that the distribution of the source masses follows the large-scale distribution of matter, indicating that our sources reside in the dense cosmic filaments.

In the bottom panels of Figure~\ref{fig:slices}, we present the $\delta T_\mathrm{b}$ slices from our `Source1\_SinkA' model. At $z\approx 5.436$, the volume-averaged neutral fraction is \xHIv{}$\approx 10^{-4}$, and so it corresponds to the final stages of reionization. We chose this epoch to understand the phase when reionization transitions from \textit{inside-out} to \textit{outside-in}, which was identified in Section~\ref{sec:reion_hist}. The bottom-left panel displays the $\delta T^\mathrm{IGM}_\mathrm{b}$, which corresponds to the signal from \hi{} in the IGM. We observe that the patterns in this signal follow the dense structures in the matter distribution (top-left panel). This correlation is caused by the higher recombination rates in the dense regions.

While most regions exhibit a very low signal strength ($\delta T^\mathrm{IGM}_\mathrm{b} \lesssim 10^{-2}\,\rm mK$), there are a few areas with high values, $\gtrsim 10\,\rm mK$. These regions correspond to neutral islands that remain shielded from UV photons. These islands are located at large enough distance from photon sources such that the recombination rate, although low, is enough to counterbalance the incoming UV photons. In the inset, we show a zoom in on one such a neutral island. 
This inset correspond to a size of $30\, \rm cMpc$ per side. 
Our simulations contain numerous such islands in other slices, and we will conduct a statistical investigation of these islands in the next subsection.

In the bottom-right panel, we show the $\delta T^\mathrm{halo}_\mathrm{b}$ produced by \hi{} inside galaxies residing in haloes, which we assigned by using the method described in Section~\ref{sec:HI_inside_haloes}. This signal follows the distribution of sources (top-right panel) and the dense cosmic filaments (top-left panel) by design. At this very late stage, the strength of this signal is comparable to the signal produced in the IGM.Within the neutral island region shown in the zoom in, the $\delta T^\mathrm{halo}_\mathrm{b}$ is predominantly low ($\lesssim 10^{-2}\,\rm mK$), where the neutral island is located. Therefore, the resulting 21-cm signal ($\delta T^\mathrm{halo}_\mathrm{b}+\delta T^\mathrm{IGM}_\mathrm{b}$) from this region will follow the under-dense matter distribution. We will discuss the impact of these islands in the statistical measurements of the 21-cm signal in subsection~\ref{sec:results_ps}.

\subsection{Neutral islands during the very late stages of reionization}\label{sec:results_islands}

In \citet{giri2019neutral}, we found few but very large (with lengths $\sim 100$ cMpc) neutral islands at \xHIv{}$\approx 10^{-2}$. With the models in this study, which extend to even later epochs ($z\lesssim 6$), we continue to observe notably large neutral islands, as discussed earlier. While Figure~\ref{fig:slices} visually presents one such large neutral island in the slice, the entire simulation volume contains more instances. We employed the mean-free-path size distribution algorithm implemented in \texttt{Tools21cm} \citep{giri2020tools21cm} to investigate these islands further. 
\RefereeReport{In this algorithm, we randomly select a neutral cell and shoot a ray in a random direction until it reaches the boundary of the neutral region. We repeat this process several times ($\sim$10$^7$) and record the lengths of the rays. The probability distribution of these lengths gives the size distribution.}
See \citet{giri2018bubble} for more information about this algorithm. 

In Figure~\ref{fig:ISDs}, we present the island size distribution (ISD), illustrating the probability of sizes for neutral islands in two sink models with the same sources (`Source1') at different epochs. The `SinkA' model is represented in blue, while the `SinkB' model is shown in red. We see that significantly large neutral islands exist in our model that can impact the statistical measurements. For the case at \xHIv{}$\approx 10^{-4}$ (dot-dashed line), `SinkA' and `SinkB' have an average value of island sizes $\bar{R} \simeq 3.2\, \rm cMpc$ and $3.6\, \rm cMpc$, respectively. 
Additionally, we observe neutral islands as large as $\sim$20 cMpc in both models.
These large islands can be detected in the upcoming image data from SKA-Low using the structure identification approaches developed in \citet{giri2018optimal} and \citet{bianco2021deep,bianco2024deep}. 
At \xHIv{}$\approx 10^{-3}$, the neutral island sizes can reach $\sim$40 cMpc.
We will discuss their impact on the power spectra in the next subsection. 

The two sink models show distinct ISDs at all the epochs in Figure~\ref{fig:ISDs}. The island sizes in the `SinkB' model is larger than that of `SinkA' at all epochs. In \citet{giri2019neutral}, we developed a model based on packing spheres in simulation volumes to comprehend the distribution of neutral islands. This model suggested that the empty spaces between spheres of large sizes are large. 
\RefereeReport{In Appendix~\ref{app:toy}, we present a modified version of this toy model that includes a distribution of sphere sizes. In scenarios with spheres of multiple sizes, the cases where the largest ionized bubbles are larger also result in neutral island size distributions biased towards larger sizes.}
Since the `SinkB' model permits the formation of larger bubbles around every source compared to `SinkA', the resulting neutral islands are more prominent, as revealed in the figure. 

\subsection{21-cm power spectrum}
\label{sec:results_ps}

\begin{figure*}
    \centering
    \includegraphics[width=1.0\linewidth]{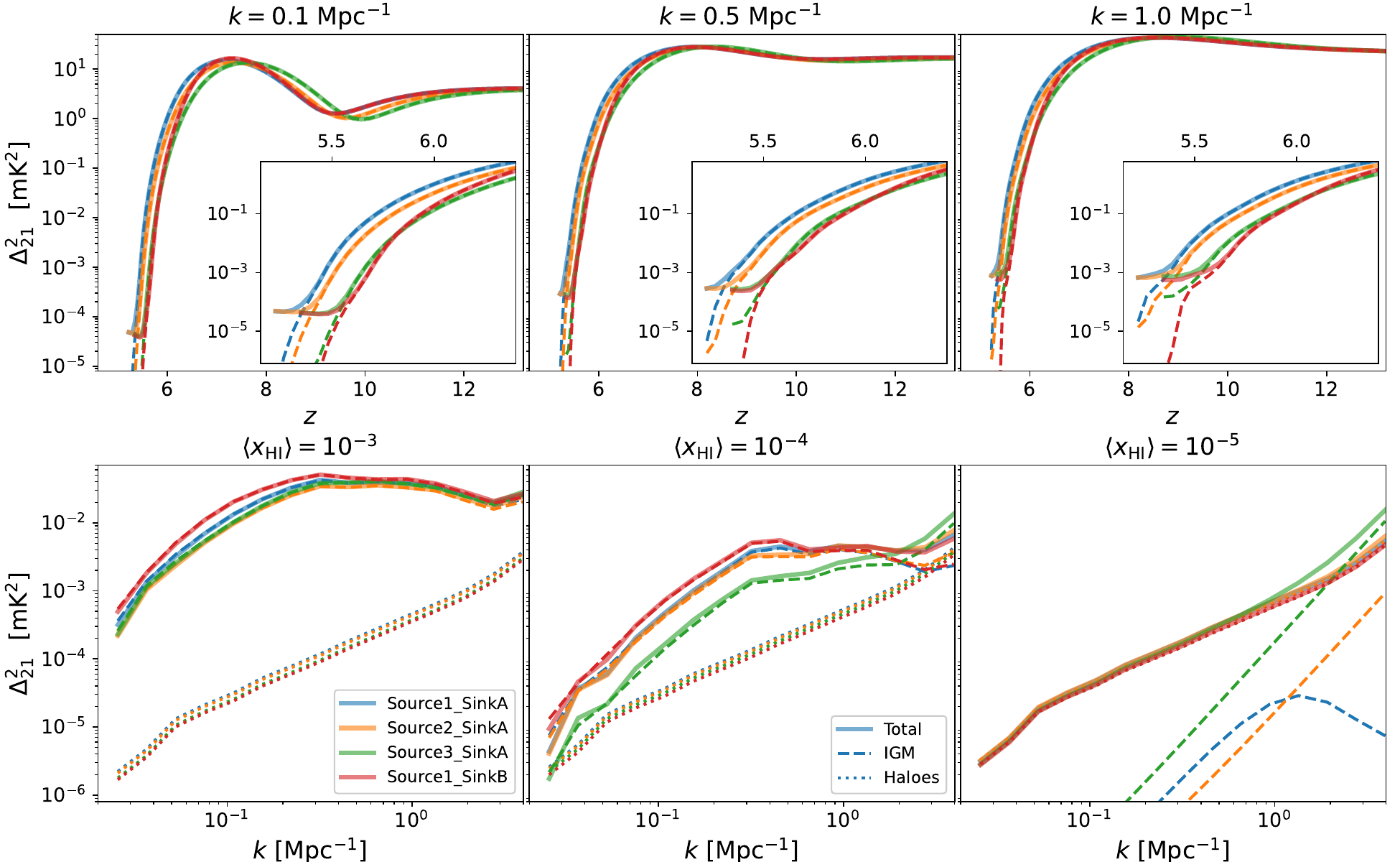}
    \caption{The 21-cm power spectra $\Delta_\mathrm{21}=k^3P_\mathrm{21}/(2\pi^2)$ for our models (Source1\_SinkA: blue, Source2\_SinkA: orange, Source3\_SinkA: green and Source1\_SinkB: red). 
    The solid lines represent $\Delta_\mathrm{21}$, assuming the contribution from \hi{} in both the IGM and inside the haloes. The dashed and dotted lines represent $\Delta_\mathrm{21}$ considering only \hi{} in the IGM and inside haloes, respectively.
    \textit{Top panel:} We show the redshift evolution of $\Delta_\mathrm{21}$ at $k=0.1$ (left), 0.5 (middle) and 1.0 (right) Mpc$^{-1}$. To focus on the end stages of reionization, the insets of the panels zoom into the region where $z\lesssim 6.5$. Until very late stages (\xHIv{}$\lesssim 10^{-4}$) of reionization, the contribution from the \hi{} in the IGM dominates.
    \textit{Bottom panel:} We compare the contribution of the \hi{} in the IGM and inside haloes to $\Delta_\mathrm{21}$ at three very late stages where \xHIv{} $\approx 10^{-3}$ (left), $10^{-4}$ (middle) and $10^{-5}$ (right). The contribution from the latter initially dominates on small scales and evolves to all scales over time.}
    \label{fig:Pk}
\end{figure*}

The spatial characteristics of the 21-cm signal can be probed with the power spectrum $(P_\mathrm{21}(k) = \langle \delta T^*_\mathrm{b}(k) \delta T_\mathrm{b}(k)\rangle)$ that the radio telescopes are attempting to detect. Post-reionization \hi{} follows the galaxy distribution and, therefore, the matter distribution. Hence, most studies model the post-reionization 21-cm power spectrum as a biased version of the matter power spectrum \citep[e.g.][]{santos2015cosmology,padmanabhan2017halo,obuljen2018high}. However, we do not make this assumption and estimate the power spectrum directly from our simulation volumes where the \hi{} both in the IGM and inside galaxies are included.

Figure~\ref{fig:Pk} illustrates the dimensionless power spectrum $\Delta_\mathrm{21}(k)=k^3P_\mathrm{21}(k)/(2\pi^2)$ of our models. The top panels show the redshift evolution at three different scales ($k\approx 0.1, 0.5, 1.0~ \mathrm{Mpc}^{-1}$ ). While the solid lines represent the $\Delta_\mathrm{21}$ of the signal produced by \hi{} from both haloes and IGM (total), the dashed lines represent the signal from the IGM alone. In these panels, we do not include the $\Delta_\mathrm{21}$ for the \hi{} inside galaxies because their magnitude is small for most redshifts. However, the noticeable difference between the solid and dashed line is caused by the inclusion of this \hi{}.

In all our models, this power spectrum is dominated by the \hi{} in the IGM until very late stages of reionization (\xHIv{}$\approx 10^{-3}$). After this epoch, the contribution from the \hi{} inside galaxies begin to become more critical. The transition of the 21-cm signal to probing the galaxies occurs when the \xHIv{} is less than $10^{-4}$. In the top panels, this epoch is marked by a flattening of the power spectra in all models and at all scales. 
By detecting this unique signature, future radio experiments would be able to provide evidence of the end of reionization. 
The epoch of the transition from IGM to galaxies coincides with the time when \xHIm{} becomes larger than \xHIv{}. As discussed in Section~\ref{sec:reion_hist}, the distribution of \hi{} in the IGM transitions from an \textit{inside-out} to \textit{outside-in} nature in our simulations.

We selected three epochs from our simulations to focus on the transition process and plot them in the bottom panels of Figure~\ref{fig:Pk}. Along with the $\Delta_\mathrm{21}$ of the total (solid lines) and IGM (dashed lines) fields, we include the field that corresponds to the \hi{} in the haloes only (dotted lines). The left panel displays the $\Delta_\mathrm{21}$ when \xHIv{} is approximately $10^{-3}$, where the power from the IGM is seen to be $\sim 2$ orders of magnitude higher than that from the galaxies. Therefore we can neglect still neglect the latter contribution at this time.

As time progresses, the small-scale signal transitions to probe the \hi{} inside haloes, as evident in the middle panel, when \xHIv{}$\approx10^{-4}$. This scale dependent transition is consistent with the findings in \citet{xu2019h}.
As discussed in Section~\ref{sec:results_islands}, a significant number of large neutral islands remains during this epoch. We observe that all models have an IGM $\Delta_\mathrm{21}$ with a knee-like feature at wave-number $k_\mathrm{knee}$ between 0.4 and 0.6 $\mathrm{Mpc}^{-1}$. This scale approximately corresponds to the average scale of the neutral islands during this time as $k_\mathrm{knee} \sim 2/\bar{R}$\footnote{In \citet{georgiev2022large}, a similar relation was identified, but with the sizes of ionized bubbles. This $k$ value determined the scale beyond which the 21-cm power spectrum became a biased tracer of matter distribution during the early stages of reionization.}. Due to this feature, we cannot assume $\Delta_\mathrm{21}$ to be a biased version of the matter power spectrum.

In the right-hand panel, we show $\Delta_\mathrm{21}$ when \xHIv{}$\approx 10^{-5}$, which is the time by when the contribution from the IGM has become negligible. Almost all the signal comes from the \hi{} inside dark matter haloes. As the $\Delta_\mathrm{21}$ is probing the haloes during this time, it can be modelled as a biased version of the matter power spectrum. Therefore, $\Delta_\mathrm{21}$ can be used to study cosmological models after this transition era.

\section{Conclusion}\label{sec:conclusion}
In this study, we enhanced our fully numerical cosmic reionization simulation framework, \pyccray{} \citepalias{hirling2024pyc}, enabling it to incorporate various high-redshift observations, including the ultraviolet luminosity functions (UVLFs) and Lyman-$\alpha$ (\lya{}) absorption spectra. Utilizing this framework, we investigated the evolution of the cosmological 21-cm signal, particularly emphasizing the very late stages of reionization. Our simulation suite was constructed based on early galaxy formation models aligned with the latest high-redshift measurements. We calibrated our model parameters for star formation within dark matter haloes 
using available UVLF measurements at high redshifts. Additionally, we considered models for UV photon escape that are consistent with constraints for the \lya{} escape fraction.

In simulation volumes such as those utilized here (with a box length of 298 cMpc), the small scale absorbers are not easily resolved. Therefore, we implemented a sink model that limits the propagation of UV photons beyond a certain distance $R_\mathrm{mfp}$. Previous studies have demonstrated the effectiveness of this sink model in explaining observations of \lya{} absorption spectra \citep[e.g.,][]{cain2021short,bosman2022hydrogen,davies2023constraints}. A comparison of several methods for modelling these sinks \RefereeReport{is presented in \citet{georgiev2024forest}.} 
We adopted two redshift dependencies for $R_\mathrm{mfp}$, calibrated to constraints from relevant studies. These models yield a distinct evolution of reionization during the later stages (\xHIv{}$\lesssim 0.5$). Most frameworks for reionization modelling typically assume a fixed $R_\mathrm{mfp}$ when interpreting observations \citep[e.g.][]{mondal2020tight,greig2021interpreting,qin2021reionization}, which may introduce biases on constraints when analyzing the final stages of reionization.

We also studied the evolution of the UV background ($\Gamma$) in our simulations, which is an output of the radiative transfer solver in \pyccray{}. This background is modulated by the sink model and, consequently, influenced by the choice of $R_\mathrm{mfp}$, consistent with findings in previous studies \citep[e.g.,][]{haardt2012radiative,sobacchi2014inhomogeneous,becker2021mean,gaikwad2023measuring}. It is important to note that although we enforce an $R_\mathrm{mfp}$ for the radiative transfer around each source, the sizes of ionized regions in our simulations can be much larger due to overlaps. Our models consistently align with constraints on $R_\mathrm{mfp}$ and $\Gamma$. This improvement is a significant enhancement compared to our previous 21-cm signal simulations \citep[e.g.][]{dixon2016large,giri2019neutral}.

A caveat of our study of the UV background is that we have considered only one method of modelling the sinks. For example, a clumping factor evolving with redshift or density-dependent clumping factor \citep[e.g.][]{mao2020impact, bianco2021impact} can affect the IGM reionization during the end stages and, consequently, affect the evolution of the $R_\mathrm{mfp}$ and $\Gamma$. Previous studies have also attributed the measured evolution of the UV background to the uniqueness of the UV emissivity during late stages \citep[e.g.][]{puchwein2019consistent,keating2020long,cain2021short}. 
\RefereeReport{\citet{cain2024rise} showed that a drop in the emissivity at $z\lesssim 6$ is necessary to explain the $\Gamma$ even after accounting for unresolved sinks. However, our simulations do not exhibit such a drop, as the redshift evolution of our emissivity is determined by the dark matter halo mass function, which grows over time. We attribute this discrepancy to differences in modeling the unresolved sinks. In the models of \citet{cain2024rise}, the sinks at these low redshifts are less efficient in absorbing photons due to photo-evaporation and pressure-smoothing during earlier times. 
}
We defer the exploration of such sink and source models to future work.

The 21-cm signal serves as a valuable probe for understanding the topology of the large-scale distribution of \hi{} in our Universe \citep{giri2021measuring,kapahtia2021prospects,elbers2023persistent}. In the post-reionization era, this signal primarily probes the \hi{} content within dark matter haloes, becoming a biased tracer of the matter power spectrum. During this period, the topology of \hi{} distribution is typically assumed to be \textit{outside-in} in nature \citep[e.g.,][]{miralda2000reionization,wyithe2008fluctuations}. In this topology, dense regions exhibit a higher neutral fraction than under-dense regions due to increased recombination rates \citep{finlator2009late,choudhury2009inside}. We investigated this transition phase and found the topology to remain \textit{inside-out} until the Universe reaches a neutral fraction of \xHIv{}$\approx 10^{-4}$, a significantly later stage compared to previous models \citep[e.g.,][]{finlator2009late}.

Our reionization models exhibit a distinct evolution of the 21-cm power spectra, highlighting the potential of this signal to constrain both source and sink models. Focusing on the very late stages of reionization (\xHIv{}$\lesssim 10^{-3}$), we studied the dependence of the signal to the \hi{} content in both the IGM and dark matter haloes. The relevance of \hi{} inside dark matter haloes becomes more pronounced after \xHIv{}$\approx 10^{-4}$, coinciding with the phase when the topology of reionization undergoes a transition in our models. The redshift evolution of the 21-cm power spectra reveals a distinct flattening at this epoch, potentially serving as a marker for the 21-cm signal entering the post-reionization era and the \hi{} topology transitioning to \textit{outside-in}.
Furthermore, we found that the transition process of the signal from probing the IGM to the haloes is a scale dependent, which begins at small-scale and transfer to large-scales. This finding is consistent with the study in \citet{xu2019h}.

In \citet{giri2019neutral}, we investigated the existence of large neutral islands during the end stages of reionization, detectable in the image data from the SKA-Low. In this study, our simulations reveal a statistically significant number of such neutral islands during the very late stages of reionization (\xHIv{}$\approx 10^{-4}$). The sizes of these islands were as large as 20 cMpc, which can be detected in the image data with our feature identification framework \citep{giri2018optimal,bianco2021deep,bianco2024deep}. 
We studied the distribution of the sizes of these islands, demonstrating that they are significant enough to impact the large-scale fluctuations of the 21-cm signal. Notably, we observe a knee-like feature in the 21-cm power spectra corresponding to these islands, persisting until the period when the signal of \hi{} in haloes starts to dominate. 

In modelling the \hi{} inside dark matter haloes, we do not consider the in- and out-flow of gas between the IGM and galaxies, as this requires sophisticated hydro-dynamical simulation. Additionally, we follow a straightforward approach using the $M_\mathrm{HI}-M_\mathrm{h}$ relation from \citet{padmanabhan2017halo} that assumes that the \hi{} in massive galaxies, residing in high mass halo $M_\mathrm{h} \gtrsim 10^{10}\,M_\odot$, can self-shield against the UV photons. However, recent \hi{} surveys provided observational evidence that suggests a more complex scenario with Ultra Faint Dwarf Galaxy able to keep most of their reservoir of \hi{} throughout EoR \citep[see][]{Irwin2007Dwarf, Saul2012GALFA, Giovanelli2013Dwarf, Janesh2019}. 
These low-mass galaxies can impact our study of the end stages of reionization, and we will explore their impact in future work.

Upcoming observations of the 21-cm signal will help study these late stages of reionization ($z\lesssim 6$) in more detail. As the low redshifts are expected to have a greater signal-to-noise ratio, we expect better constraining capability during these stages. Therefore, our models can be helpful in interpreting the signal during these late epochs. 

\section*{Acknowledgements}
We acknowledge Benoit Semelin, Joachim Stadel, and Yves Revaz for their helpful discussions. 
\RefereeReport{We also thank the anonymous reviewer for their very helpful comments.}
Nordita is supported in part by NordForsk. GM is supported by the Swedish Research Council project grant 2020-04691\_VR and AS is supported by the Swiss National Science Foundation (SNF) via the grant PCEFP2\_181157.
We acknowledge the allocation of computing resources provided by the National Academic Infrastructure for Supercomputing in Sweden (NAISS) and the Swiss National Supercomputing Centre (CSCS) under the SKA share with the project ID sk015. We have utilised the following \texttt{Python} packages for manipulating the simulation outputs and plotting results: {\tt numpy} \citep{van2011numpy}, {\tt scipy} \citep{virtanen2020scipy}, and {\tt matplotlib} \citep{hunter2007matplotlib}. 

\section*{Data Availability}
The source code used
for the simulations of this study, the \pkdgrav{} (\url{https://bitbucket.org/dpotter/pkdgrav3/}) and \pyccray{} (\url{https://github.com/cosmic-reionization/pyC2Ray}), are publicly available. We have made the 21-cm signal simulation data freely available at \url{https://doi.org/10.5281/zenodo.10785609}. 
Our data can be read and manipulated using our public tool, {\tt Tools21cm} (\url{https://github.com/sambit-giri/tools21cm}).




\bibliographystyle{mnras}
\bibliography{refs} 

\appendix

\section{Validating the halo catalogue}\label{app:hmf}
Our simulation framework, \pyccray{}, utilizes the dark matter haloes to model photon sources in our cosmological simulations. In this study, we have used the friend-of-friends algorithm implemented in \pkdgrav{} to find dark matter haloes in snapshots from \nbody{} simulation. As we wanted to model reionization at large scales ($\gtrsim$100 Mpc) caused by sources residing in haloes as small as $\sim$10$^9 M_\odot$, we had to find a balance between the box size and the number of particles. To ensure this halo catalogue, we compare the corresponding halo mass function (HMF) against the ones modelled with extended Press-Schechter (EPS) calculation. 

\begin{figure}
    \centering
    \includegraphics[width=1.0\linewidth]{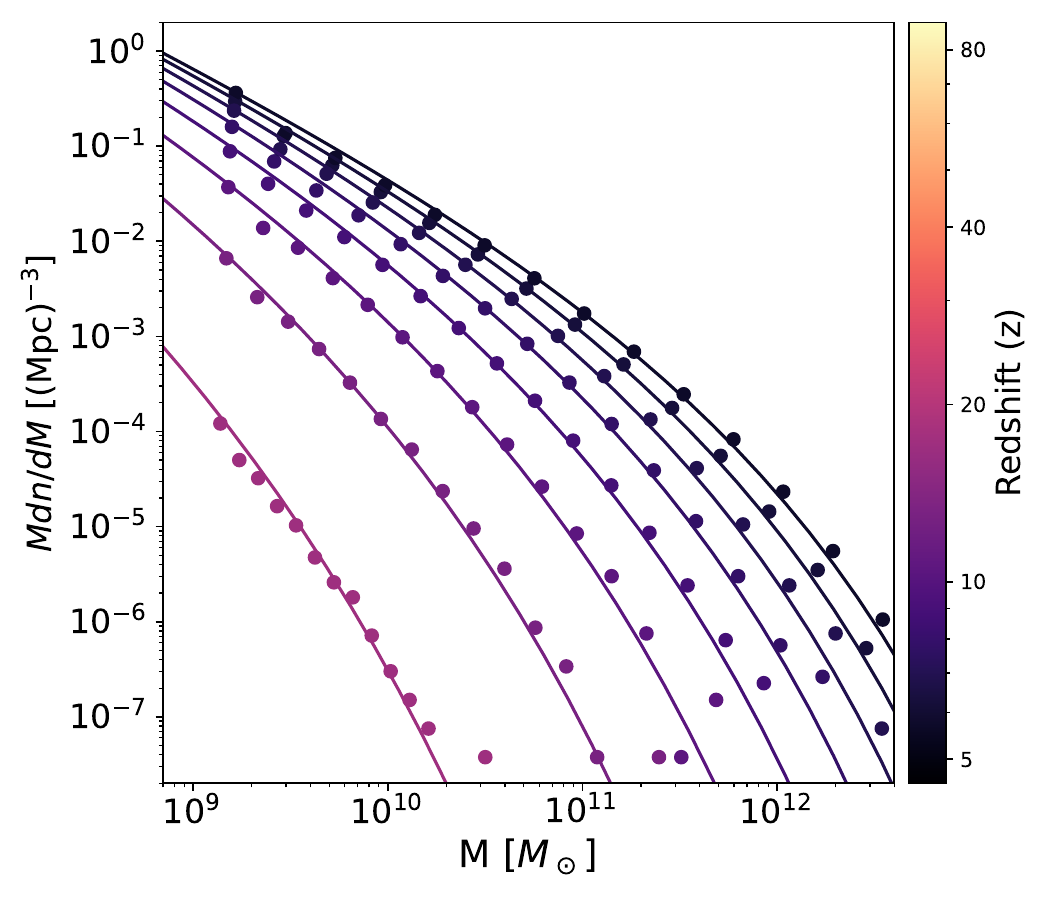}
    \caption{The halo mass function calculated from the halo catalogue of our \nbody{} simulations (points) and modelled using extended Press-Schechter formalism (solid lines). The redshift values are marked with the colour of the lines.}
    \label{fig:hmf_eps}
\end{figure}

We follow the EPS formalism described in \citet{Schneider2018constraining} and refer interested readers to this work and the references therein. In Figure~\ref{fig:hmf_eps}, we observe a generally robust agreement among the HMFs across all redshifts. However, some discrepancies are noticeable, particularly at the high-mass end. This discrepancy is attributed to the substantial Poisson noise inherent in modelling the massive haloes within the selected simulation volume.

\RefereeReport{
\section{Toy Models for neutral islands}
\label{app:toy}
}
\RefereeReport{
We developed a modified version of the toy model presented in \citet{giri2019neutral} to understand the impact of sink models on the sizes of neutral islands. This model was based on packing spheres or ionized bubbles into a simulation volume. The box length and resolution of these toy models were the same as those of the cosmological reionization simulations presented in this work. Previous authors \citep[e.g.,][]{Bharadwaj:2004nr,zaldarriaga200421,majumdar2018quantifying} have used similar approaches to study the impact of ionized bubbles on the spatial fluctuations in the 21-cm signal.
In this work, we uniformly sampled a list of sizes up to a maximum radius of $R_\mathrm{max}$ and painted the corresponding spheres in the simulation volume. We continued this process until achieving a certain volume filling fraction ($f_\mathrm{v}$) of the bubbles.
}

\begin{figure}
    \centering
    \includegraphics[width=0.96\linewidth]{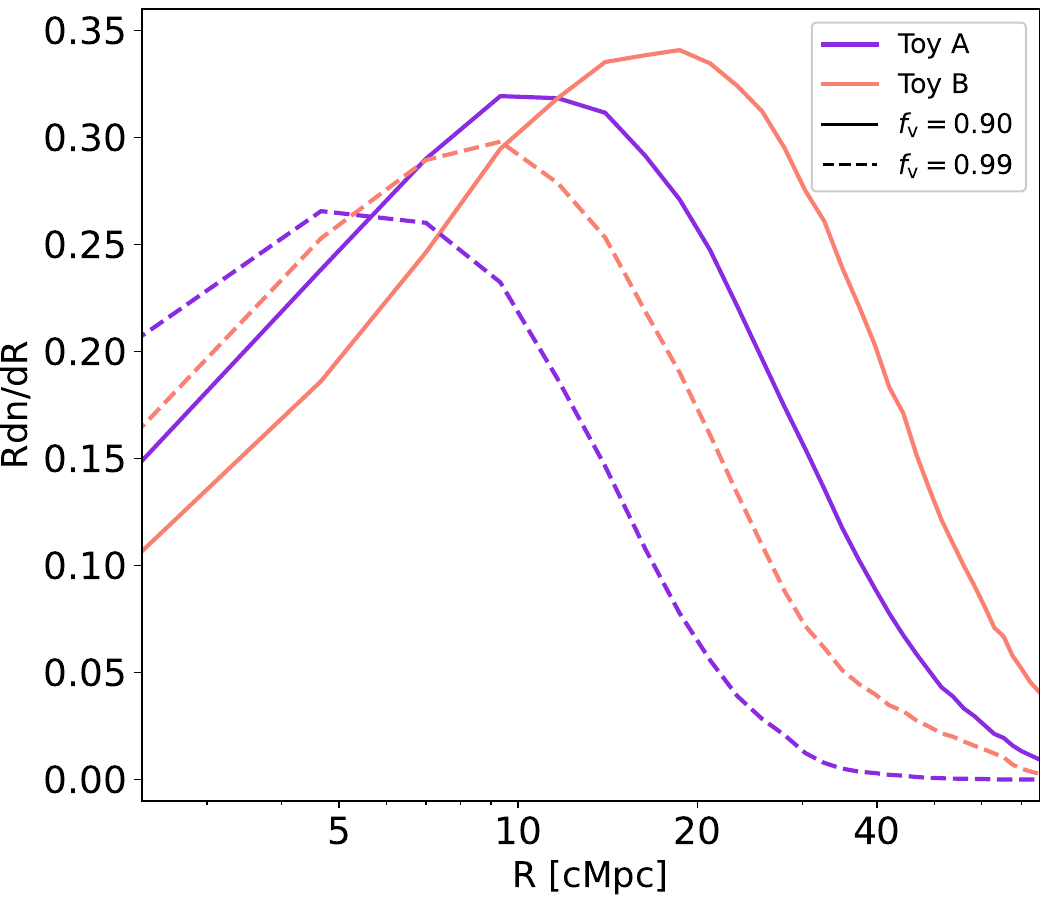}
    \caption{\RefereeReport{The size distribution of neutral islands in two toy cases at two bubble filling fractions ($f_\mathrm{v}$). Toy A and Toy B contain ionized bubbles with a maximum radius of 10 and 20 Mpc, respectively. The neutral islands in Toy B are larger compared to those in Toy A.}}
    \label{fig:toy_bubble}
\end{figure}

\RefereeReport{
Using this approach, we produced Toy A and Toy B models with $R_\mathrm{max}$ set to 10 and 20 Mpc, respectively. In Figure~\ref{fig:toy_bubble}, we show the size distribution of the neutral islands in both toy models at $f_\mathrm{v}=0.90$ (solid lines) and 0.99 (dashed lines). We observe that the sizes are larger in Toy B compared to Toy A. In \citet{giri2019neutral}, we showed that large bubbles result in large neutral regions. Unlike the toy model in \citet{giri2019neutral}, we included bubbles of smaller sizes, which could fill the large neutral cavities created by the largest bubbles. However, in this modified model, we observe that the sizes of neutral islands are sensitive to the largest bubble sizes in the volume. Even though the small bubbles have the capacity to break up the large neutral regions, Toy B contains enough neutral cavities that were unaffected, resulting in most probable island sizes larger than those in Toy A. We should note that we are comparing the scenarios at the same volume fraction and, therefore, the number of bubbles in Toy B could be fewer than in Toy A.
}

\bsp	
\label{lastpage}
\end{document}